\documentclass[10pt,conference]{IEEEtran}
\IEEEoverridecommandlockouts

\usepackage{array,multirow,etoolbox,graphicx,subcaption,fancyhdr,tabularx,longtable}
\usepackage{float}
\usepackage{graphicx}
\usepackage{amsmath} 
\usepackage{tcolorbox}
\usepackage{booktabs}
\usepackage{enumitem}
\usepackage{mathtools}
\usepackage{xcolor}
\usepackage{hyperref}
\usepackage{balance}

\def\baselinestretch{1.02}

\newtcolorbox{TcolorBox}[1]{fonttitle=\bfseries,title=#1}

\usepackage{xspace}
\usepackage{dirtytalk}
\usepackage{xcolor}

\definecolor{teal}{RGB}{0,128,128}

\newcommand{\Jiyang}[1]{\textcolor{purple}{{\it [Jiyang says: #1]}}}

\newcommand{\etal}{\emph{et al.}\xspace}

\newcommand{\DefMacro}[2]{\expandafter\newcommand\csname rmk-#1\endcsname{#2}}
\newcommand{\UseMacro}[1]{\csname rmk-#1\endcsname}
\newcommand{\rom}[1]{\uppercase\expandafter{\romannumeral #1\relax}}

\newcommand{\Title}{Using Large-scale Heterogeneous Graph Representation Learning for Code Review Recommendations at Microsoft}

\newcommand{\pr}{pull request}

\newcommand{\kg}{Socio-technical graph}
\newcommand{\NumRepos}{332} 
\newcommand{\NumNodes}{5,858,834} 
\newcommand{\NumEdges}{23,803,053} 
\newcommand{\NegFeedbackCats}{3}
\newcommand{\EvalPullRequest}{254K}
\newcommand{\UserStudySize}{500}

\newcommand{\nrr}{\textsc{Coral}}
\newcommand{\NRR}{\textsc{Coral}\xspace}
\newcommand{\hm}{rule-based model}
\newcommand{\HM}{Rule-based Model\xspace}
\newcommand{\PRstartYear}{2019}

\newcommand{\anecdote}[1]{\say{\emph{#1}}}
\newcommand{\NalandaGraph}{socio-technical graph}
\newcommand{\engagementpct}{precision\xspace}

\newcommand{\MyPara}[1]{\noindent\textbf{#1}}

\newcommand{\TableCaptionUserResponse}{Distribution of qualitative user study responses. \label{tab:UserResponse}}
\newcommand{\TableCaptionNegativeFeedback}{Users' Negative Feedback Categories. \label{tab:NegFeedback}}
\newcommand{\TableCaptionUserStudyResults}{Comparative user study \engagementpct across dimensions. RM is Rule-based Model. The differences between the two models with the same Greek letter suffix (and only those pairs) are not statistically significant. \label{tab:UserStudy}}

\DefMacro{UserDataSampleSize}{500}

\DefMacro{YoungSmallBaselineAcc}{0.38} 
\DefMacro{YoungMediumBaselineAcc}{0.36}  
\DefMacro{YoungLargeBaselineAcc}{0.25}
\DefMacro{MediumSmallBaselineAcc}{0.32}
\DefMacro{MediumMediumBaselineAcc}{0.34}  
\DefMacro{MediumLargeBaselineAcc}{0.21}  
\DefMacro{OldLargeBaselineAcc}{0.16}  
\DefMacro{OldMediumBaselineAcc}{0.26}  
\DefMacro{OldSmallBaselineAcc}{0.42}  

\DefMacro{YoungSmallNRRAcc}{0.38}
\DefMacro{YoungMediumNRRAcc}{0.57}
\DefMacro{YoungLargeNRRAcc}{0.50}
\DefMacro{MediumSmallNRRAcc}{0.21}
\DefMacro{MediumMediumNRRAcc}{0.33}
\DefMacro{MediumLargeNRRAcc}{0.39}
\DefMacro{OldLargeNRRAcc}{0.34}
\DefMacro{OldMediumNRRAcc}{0.31}
\DefMacro{OldSmallNRRAcc}{0.21}

\DefMacro{NumberResponseA}{93}
\DefMacro{RateResponseA}{32.40\%}  
\DefMacro{NumberResponseB}{24}
\DefMacro{RateResponseB}{8.36\%}  
\DefMacro{NumberResponseC}{170}
\DefMacro{RateResponseC}{59.23\%}  
\DefMacro{NumberTotalResponse}{287}
\DefMacro{RatePositiveResponse}{67.6\%}
\DefMacro{RateNegativeResponse}{32.4\%}

\newcommand{\feedback}[1]{\rom{#1}}

\DefMacro{NumberFeedback1}{54}
\DefMacro{RateFeedback1}{69.23\%}  
\DefMacro{NumberFeedback2}{17}
\DefMacro{RateFeedback2}{21.79\%}  
\DefMacro{NumberFeedback3}{5}
\DefMacro{RateFeedback3}{6.41\%}  
\DefMacro{NumberFeedback4}{2}
\DefMacro{RateFeedback4}{2.56\%}  
\DefMacro{NumberFeedbackTotal}{78}
\DefMacro{NumberFeedbackCategory1Combined}{71}
\DefMacro{RateFeedbackCategory1Combined}{91.03\%}

\DefMacro{SmallBaselineAcc}{0.35} 
\DefMacro{MediumBaselineAcc}{0.31}
\DefMacro{LargeBaselineAcc}{0.19}

\DefMacro{SmallNRRAcc}{0.23}
\DefMacro{MediumNRRAcc}{0.36}
\DefMacro{LargeNRRAcc}{0.37}

\DefMacro{Top1RecoveryPct}{1.90}
\DefMacro{Top1RecoveryNum}{97}
\DefMacro{Top3RecoveryPct}{4.50}
\DefMacro{Top3RecoveryNum}{223}
\DefMacro{Top5RecoveryPct}{7.30}
\DefMacro{Top5RecoveryNum}{357}
\DefMacro{Top7RecoveryPct}{8.80}
\DefMacro{Top7RecoveryNum}{432}

\newcommand{\RQone}{How well does \NRR model the review history?}

\begin{document}
\title{\Title{}}

\author{\IEEEauthorblockN{Jiyang Zhang*}
\IEEEauthorblockA{\textit{The University of Texas at Austin} \\
jiyang.zhang@utexas.edu}
\\
\IEEEauthorblockN{Christian Bird}
\IEEEauthorblockA{\textit{Microsoft Research} \\
cbird@microsoft.com}
\\
\IEEEauthorblockN{Yamini Jhawar}
\IEEEauthorblockA{\textit{Microsoft  Research} \\
t-yajhawar@microsoft.com}
\and
\IEEEauthorblockN{Chandra Maddila*}
\IEEEauthorblockA{\textit{Microsoft Research} \\
maddilac@acm.org}
\\
\IEEEauthorblockN{Ujjwal Raizada }
\IEEEauthorblockA{\textit{Microsoft  Research} \\
ujjwalraizada@gmail.com}
\\
\IEEEauthorblockN{Kim Herzig}
\IEEEauthorblockA{\textit{Microsoft} \\
kimh@microsoft.com}
\and
\IEEEauthorblockN{Ram Bairi}
\IEEEauthorblockA{\textit{Microsoft Research} \\
rbairi@microsoft.com}
\\
\IEEEauthorblockN{Apoorva Agrawal}
\IEEEauthorblockA{\textit{Microsoft Research} \\
t-aagraw@microsoft.com}
\\
\IEEEauthorblockN{Arie \lowercase{van} Deursen}
\IEEEauthorblockA{\textit{Delft University of Technology} \\
Arie.vanDeursen@tudelft.nl}

}

\maketitle
\thispagestyle{plain}
\pagestyle{plain}
\def\thefootnote{*}\footnotetext{Work performed while at Microsoft Research; Equal contribution}


\begin{abstract}

Code review is an integral part of any mature software development process, and identifying the best reviewer for a code change is a well-accepted problem within the software engineering community.
Selecting a reviewer who lacks expertise and understanding can slow development or result in more defects.
To date, most reviewer recommendation systems rely primarily on historical file change and review information; those who changed or reviewed a file in the past are the best positioned to review in the future.

We posit that while these approaches are able to identify and suggest qualified reviewers, they may be blind to reviewers who have the needed expertise and have simply never interacted with the changed files before.
Fortunately, at Microsoft, we have a wealth of work artifacts across many repositories that can yield valuable information about our developers.
To address the aforementioned problem, we present \NRR, a novel approach to reviewer recommendation that leverages a \NalandaGraph{} built from the rich set of entities (developers, repositories, files, pull requests (PRs), work items, etc.) and their relationships in modern source code management systems.
We employ a graph convolutional neural network on this graph and train it on two and a half years of history on \NumRepos{} repositories within Microsoft.

We show that \NRR is able to model the manual history of reviewer selection remarkably well.
Further, based on an extensive user study, we demonstrate that this approach identifies relevant and qualified reviewers who traditional reviewer recommenders miss, and that these developers desire to be included in the review process.
Finally, we find that ``classical'' reviewer recommendation systems perform better on smaller (in terms of developers) software projects while \NRR excels on larger projects, suggesting that there is ``no one model to rule them all.''

\end{abstract}

\section{Introduction} 

Code review (also known as pull request review) has become an integral process in software development, both in industrial and open source development~\cite{gousios2014exploratory,rigby2012contemporary,rigby2013convergent} and all code hosting systems support it.
Code reviews facilitate knowledge transfer, help to identify potential issues in code, and promote discussion of alternative solutions~\cite{bacchelli2013expectations}.
Modern code review is characterized by asynchronous review of changes to the software system, facilitated by automated tools and infrastructure~\cite{bacchelli2013expectations} .

As code review inherently requires expertise and prior knowledge, many studies have noted the importance of identifying the ``right'' reviewers, which can lead to faster turnaround, more useful feedback, and ultimately higher code quality~\cite{rigby2011understanding,bosu2015characteristics}.
Selecting the wrong reviewer slows down development at best and can lead to post-deployment issues.
In response to this finding, a vibrant line of code reviewer recommendation research has emerged, to great success~\cite{lipcak2018large,ouni2016search,jiang2017should,yu2014reviewer,yu2016reviewer,lee2013patch,sulun2019reviewer,thongtanunam2015should}.
Some of these have, in fact, even been put into practice in industry~\cite{asthana2019whodo}.

All reviewer recommender approaches that we are aware of rely on historical information of changes and reviews. The principle underlying these is that best reviewers of a change are those who have previously authored or reviewed the files involved in the review.
While recommenders that leverage this idea have proven to be valid and successful, we posit that they may be blind to qualified reviewers who may have never interacted with these files in the past, especially as the number of developers in a project grows.


We note that there is a wealth of additional recorded information in software repositories that can be leveraged to improve reviewer recommendation and address this weakness.
Specifically, we assert that incorporating information around interactions between code contributors as well as the semantics of code changes and their descriptions can help identify the best reviewers.
As one intuitive example, if a set of existing pull requests (PRs) are determined to be semantically similar to a new incoming pull request, then reviewers who contributed meaningfully to the former may likely be good candidates to review the latter, even if the reviews do not share common files or components.
To leverage this idea, we construct a \NalandaGraph{} on top of the repository information, comprising files, authors, reviewers, pull requests, and work items, along with the relationships that connect them.
Prior work has shown that code review is a social process in addition to a technical one~\cite{kononenko2015investigating,bosu2013impact}.
As such, our primary belief is that this heterogeneous graph captures both and can be used to address various software engineering tasks, with code reviewer recommendation being the first that we address.

Learning on such a graph poses a challenge.  Fortunately, the area of machine learning has advanced by leaps and bounds in the past years since reviewer recommendation became a recognized important research problem.
Neural approaches give us tools to deal with this relational information found in software repositories and make inferences about who is best able to review a change~\cite{wu2020comprehensive}.

Based on these observations and ideas, we introduce \NRR, a novel approach for identifying the best reviewers for a code change.  
We train a graph convolutional neural network on this socio-technical graph and use it to recommend reviewers for future pull requests.
Compared to existing state of the art, this approach works quite well.

To test our hypotheses, we build a  \NalandaGraph{} of the entities and their relationships in \NumRepos{}
software projects over a two and a half year period at Microsoft.
We show that a neural network trained on this graph is able to model review history surprisingly well.  
We perform a large scale user study of \NRR by contacting those potential reviewers recommended by our neural approach that the ``classical'' baseline (in production) approach did not identify because they did not previously interact with the files in the pull requests.
Their responses reveal that there is a large population of developers who not only are qualified to contribute in these code reviews, but that desire to be involved as well.
We also investigate in what contexts \NRR works best and find it performs better than the baseline in large (in terms of developers) software projects, but the baseline excels in small projects, indicating that there is ``no one model to rule them all.''
Finally, through an ablation study of \NRR, we demonstrate that while both files and their natural language text in the graph are important, there is a tremendous performance boost when used together.

We make the following contributions in this paper:

1. We present a general \NalandaGraph{} based on the entities and interactions in modern source code repository systems.

2. We introduce \NRR, a novel code reviewer recommendation approach that leverages graph convolutional neural networks on the socio-technical repository graph.

3. We evaluate our approach through retrospective analyses, a large scale user study, and an ablation study to show that \NRR improves on the state of the art deployed approaches on a broad scale of historical reviews and also conduct a user study based running our system on real-time PRs.

\section{Related work}

There have been many approaches to the code reviewer recommendation problem. 
We survey a broad set of studies and approaches here and refer the reader to the work of {\c{C}}etin~\etal~\cite{ccetin2021review} for a more comprehensive survey of existing work.

The first reviewer recommendation system we are aware of was introduced by Balachandran~\etal~\cite{balachandran2013reducing}.
They used authorship of the changed lines in a code review (using \texttt{git blame}) to identify who had worked on that code before and suggested a ranked list of this set as potential reviewers.
Lipcak and Rossi~\cite{lipcak2018large} performed a large scale (293,000 pull requests) study of reviewer recommendation systems.  They found that no single recommender works best for all projects, further supporting our assertion that there is no ``one recommender to rule them all.''
Thongtanunam~\etal~\cite{thongtanunam2015should} proposed \textsc{RevFinder}, a reviewer recommender based on file locations.  \textsc{RevFinder} is able to recommend reviewers for new files based on reviews of files that have similar paths in the filesystem.  The approach was evaluated on over 40,000 code reviews across three OSS projects, and recalls a correct reviewer in the top 10 recommendations 79\% of the time on average.
S{\"u}l{\"u}n~\etal~\cite{sulun2019reviewer} construct an artifact graph similar to our socio-technical graph and recommend potential reviewers based on paths through this graph from people to the artifact under review.
Lee~\etal~\cite{lee2013patch} build a graph of developers and files with edges indicating a developer committed to a file or one file is close to another file in the Java namespace tree.  They use a random walk approach on this graph to recommend reviewers.

Yu~\etal~\cite{yu2014reviewer,yu2016reviewer} recommend reviewers for a pull request by examining other pull requests whose terms have high textual similarity (cosine similarity in a term vector space), the \emph{comment network} of other developers who have commented on the author's pull requests in the past, and prior social interactions of developers with the author on GitHub.
Jiang~\etal~\cite{jiang2017should} examine the impact of various attributes of a pull request on a reviewer recommender, including file similarity, PR text similarity, social relations, and ``activeness'' and time difference.
They find that adding measures of activeness to prior models increases performance considerably.
Ouni~\etal~\cite{ouni2016search} used a genetic search-based approach to find the optimal set of reviewers based on their \emph{expertise} on the files involved and their previous \emph{collaboration} with the change author.
Zanjani~\etal~\cite{zanjani2015automatically} train a model of expertise based on author interactions with files and a time decay to provide ranked lists of potential reviewers for a given change set.
Rahman~\etal~\cite{rahman2016correct} propose \textsc{CORRECT} an approach to recommend reviewers based on their history across all of GitHub as well as their experience certain specialized technologies associated with a pull request.

Dougan \etal~\cite{dougan2019investigating} investigated the problem of ground truth in reviewer recommendation systems.
They point out that many tools are trained and evaluated on historical code reviews and rely on an (often unstated) assumption that the selected reviewers were the correct reviewers.  
They find that using history as the ground truth is inherently flawed.

\section{\NalandaGraph{}} \label{graph}

\begin{figure}
\centering
\includegraphics[width=1\columnwidth]{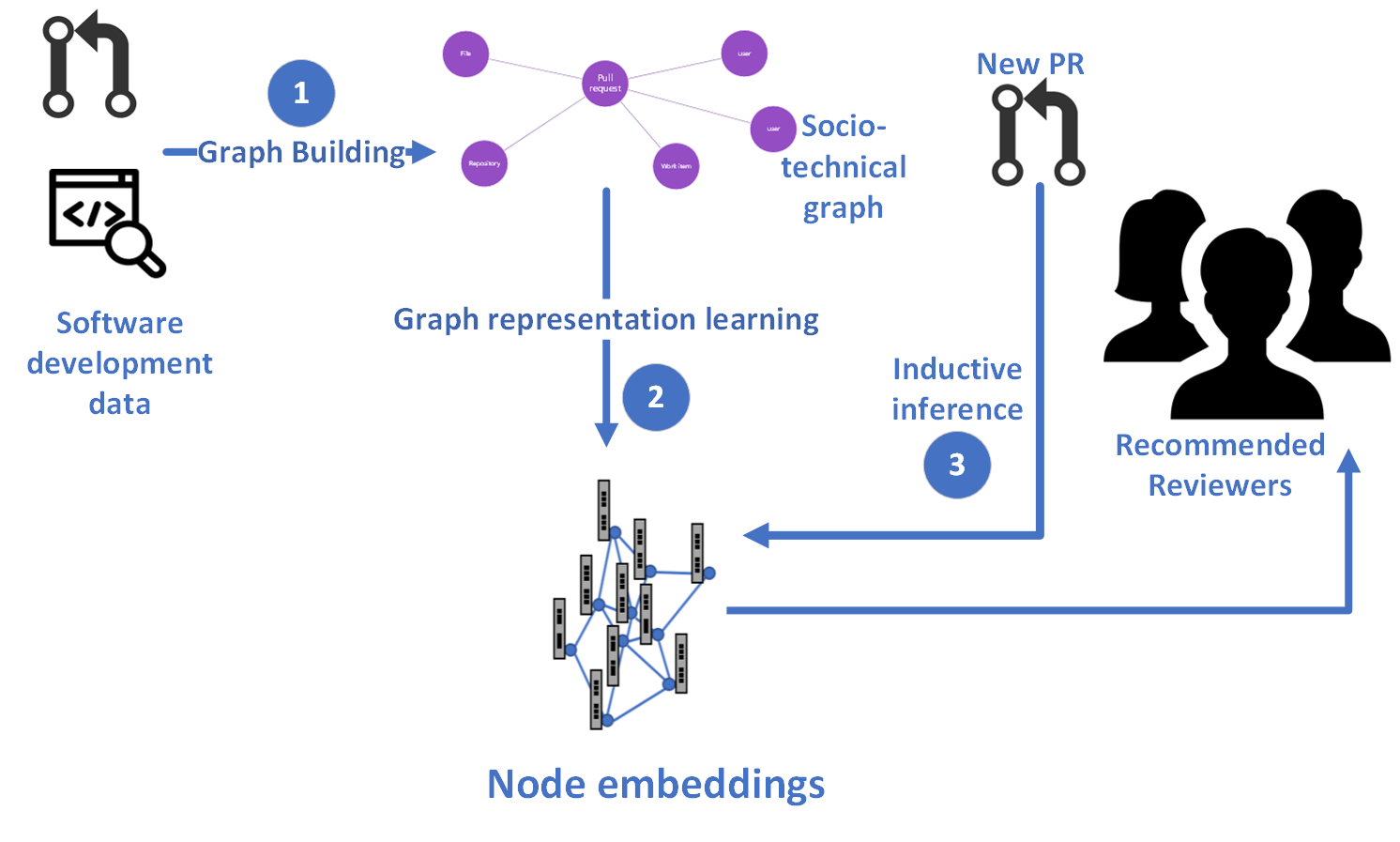}
\caption{\NRR architecture}
\label{fig:Architecture}
\end{figure}



\nrr{} system contains three main building blocks (as shown in Figure \ref{fig:Architecture}). 
\begin{enumerate}
    \item Building the \NalandaGraph{}.
    \item Performing the graph representation learning to generate node embeddings.
    \item Performing inductive inference to predict reviewers for the new \pr{}s.
\end{enumerate}

In this section, we describe the process of building the \NalandaGraph{} from entities (developers, repositories, files, pull requests, work items) and their relationships in modern source code management systems shown as step 1 in Figure \ref{fig:Architecture}.

\subsection{Socio-technical Graph} \label{stg}
\begin{figure}
\centering
\includegraphics[width=1\columnwidth]{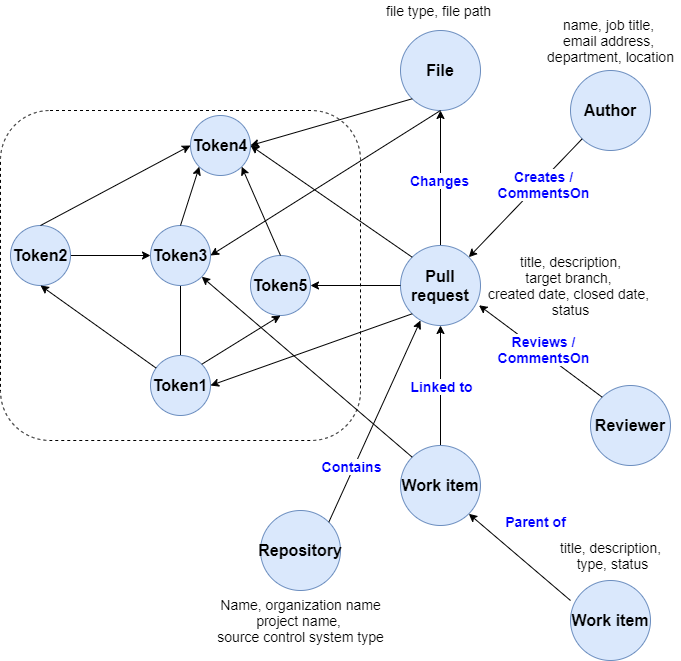}
\caption{\kg{}}
\label{fig:ASTKG}
\end{figure}

The \kg{} consists of nodes, which represent the people and the artifacts,  and edges, which represent the relationships or interactions that exist between the nodes. Figure \ref{fig:ASTKG} shows the nodes and the edges along with their properties. The \kg{} (STG) has two fundamental elements. \\
\MyPara{Nodes} There are six types of nodes in the \kg{}. They are \pr{}, work item, author, reviewer, file, and repository. \\
\MyPara{Edges} There are five types of edges in the \kg{} as listed below. \\
\texttt{creates} created between an author node and a \pr{} node. \\
\texttt{reviews} created between a reviewer node and a \pr{} node. \\
\texttt{contains} created  between a repository node and a \pr{} node if the repository contains the \pr{}. \\
\texttt{changes} created between a \pr{} node and a file node if the \pr{} is changing the file. \\ 
\texttt{linkedTo} created between a \pr{} node and a work item node if the \pr{} is linked to the work item. \\
\texttt{commentsOn} created between a \pr{} node and a reviewer node if the reviewer places code review comments. \\
\texttt{parentOf} created between a work item node and another work item node if there exist a parent-child relationship between them. \\

\subsection{Augmented \kg} \label{augmented-stg}
To include semantic information, we expand the \kg{} to have text tokens represented as nodes. This has two benefits: 
\begin{enumerate}
    \item Map users to concepts (word tokens): this helps in building a knowledge base of users (authors, reviewers) to concepts. For example., if a user is authoring/reviewing \pr{}s which contain a token, a second order relationship will be established from that user to the token.
    \item Bring semantically similar token together: as we are establishing edges between words that appear together, we capture the semantic similarity between the words. 
\end{enumerate}
We perform the four steps explained below to construct the augmented \kg{} (ASTG):
\begin{itemize}
    \item Tokenize the text (title and description) of each \pr{}, work item, and the names of the source code files edited in those \pr{}s. \item Filter the stop words by implementing a block list \cite{StopWords}.
    \item All the text nodes that appear in a \pr{} title or description and work item title or description are linked to the respective \pr{}s. All the text nodes that appear in a file name are linked to the file nodes.
    \item Text nodes are linked to each other based on their co-occurrence in the \pr{} corpus. Pointwise Mutual Information (PMI) \cite{manning99foundations} is a common measure of the strength of association between two terms.
\begin{equation}
 PMI(x,y) = \log \frac{p(x,y)} {p(x)p(y)} \label{eq:special}
\end{equation}

The formula is based on maximum likelihood estimates: when we know the number of observations for token x, o(x), the number of observations for token y, o(y) and the size of the corpus N, the probabilities for the tokens x and y, and for the co-occurrence of x and y are calculated by:
\begin{equation}
p(x) = \frac{o(x)} {N} \quad
p(y) = \frac{o(y)} {N} \quad
p(x,y) = \frac{o(x,y)} {N}
\end{equation}
The term p(x,y) is the probability of observing x and y together. \\

\end{itemize}

\begin{table}
\centering
\caption{Distribution of node and edge types in the \kg}
\begin{tabular}{llr}
\toprule
Element type & Label & Count \\
\midrule
Node  & \pr{} & 1,342,821  \\
Node  & work item & 542,866  \\
Node  & file & 2,809,805  \\
Node  & author & 18,001  \\
Node  & reviewer & 30,585  \\
Node  & text & 1,104,427  \\ 
\midrule
Total & & \NumNodes \\
\midrule
Edge  & creates & 1,342,821  \\
Edge  & reviews & 7,066,703  \\
Edge  & contains & 1,342,821  \\
Edge  & changes & 12,595,859  \\
Edge  & parent of & 148,422  \\
Edge  & linked to & 1,252,901  \\
Edge  & comments on & 53,506  \\
\midrule
Total & & \NumEdges\ \\
\bottomrule
\end{tabular}
\label{tab:NodeEdgeDistribution}
\end{table}

\subsection{Scale} \label{stg-sclae}
The \kg\ is built using the software development activity data from \NumRepos\ repositories. We ingest data starting from 1\textsuperscript{st} January, 2019, or from when the first \pr{} is created in a repository (whichever is older). The graph is refreshed three times a day. During the refresh we perform two operations: \\
\MyPara{Insert} Ingest new \pr{}s, work items, and code review information, across all the \NumRepos\ repositories, by creating corresponding nodes, edges, properties. \\
\MyPara{Update} the word tokens connected to nodes, if there are changes. We also update the edges between nodes to reflect the changes in the source data.

The \kg\ contains \NumNodes\ nodes and \NumEdges\ edges. A detailed statistics of node and edge types can be found in Table \ref{tab:NodeEdgeDistribution}.

\section{Reviewer Recommendation via Graph Neural Networks}

Reviewing a pull request is a collaborative effort. Good reviewers
are expected to write good code review comments that help improve
the quality of the code and thus shape a good product. In order to
achieve this, a good reviewer needs to be 1) familiar with the feature
that is implemented in the pull request, 2) experienced in working
with the source code and the files that are modified by the pull request,
3) a good collaborator with others in the team, and, 4) actively involved
in creating and reviewing related pull requests in the repository.
Hence a machine learning algorithm that recommends reviewers for a
pull request needs to model these complex interaction patterns to
produce a good recommendation. Feature learning via embedding generation
has shown good promise in the literature in capturing complex patterns
from the data \cite{hoff2002latent,hamilton2017representation,chen2018tutorial,zhou2021graph}.
Hence in this work we propose to pose the reviewer recommendation
problem as ranking reviewers using similarity scores between the users
and the pull requests in the \textit{embedding space}. In the rest
of this section we give details on learning embedding for pull requests
and users along with other entities (such as files, word tokens, etc.), and scoring
top reviewers for a new pull request using the learned embeddings.

The \NalandaGraph{} shown in Figure \ref{fig:ASTKG} has essential ingredients
to model the characteristics of a good reviewer: 1) the \textit{user
- pull request - token} path in the graph associates a user to a set
of words that characterize the user's familiarity with one or more
topics. 2) \textit{user - pull request - file} path associates a user
to a set of files that the user authors or reviews. 3) \textit{user
- pull request - user} path characterizes the collaboration between
people in a project. 4) \textit{pull request - user - pull request}
path characterizes users working on related pull requests. Essentially,
by envisioning software development activity as an interaction graph
of various entities, we are able to capture interesting and complex
relations and patterns in the system. We aim to encode all these complex
interactions into entity embeddings using Graph Neural Network (GNN)
\cite{wu2020comprehensive}. These embeddings are then used as features to predict most relevant reviewers to a pull request. In Figure \ref{fig:Architecture} this is depicted as step 2 and 3.

\subsection{Graph Neural Network Architecture}

Graph Convolutional Network (GCN) \cite{schlichtkrull2017modeling}
(which is a form of GNN) has shown great success in the machine learning
community for capturing complex relations and interaction patterns
in a graph through node embedding learning. In GCN, for each node,
we aggregate the feature information from all its neighbors and of
course, the feature itself. During this aggregation, neighbors are
weighted as per the edge (relation) weight. A common approach that
has been used effectively in the literature is to weigh the edges
using a symmetric-normalization approach. Here we normalize the edge
weight by the degrees of both the nodes connected by the edge. The
aggregated feature values are then transformed and fed to the next
layer. This procedure is repeated for every node in the graph. 

Mathematically it can be represented as follows:

\begin{equation}
h_{u}^{\left(k\right)}=\sigma\left(\sum_{v\in\mathcal{N}\left(u\right)\cup\left\{ u\right\} }\frac{W^{\left(k-1\right)}h_{v}^{\left(k-1\right)}}{\sqrt{\left|\mathcal{N}\left(u\right)\right|\left|\mathcal{N}\left(v\right)\right|}}\right)\label{eq:gcn}
\end{equation}

where $h_{u}^{\left(k\right)}$ is the embedding of node $u$ in the
$k^{\text{th}}$ layer; $h^{\left(0\right)}$ is the initial set of
node features, which can be set to one-hot vectors if no other features
are available; $\mathcal{N}\left(u\right)$ is the set of neighbors
of node $u$; $W^{\left(k\right)}$ is the feature transformation
weights for the $k^{\text{th}}$step (learned via training), $\sigma$
is the activation function (such as RELU \cite{relu}). Note that
symmetric-normalization is achieved by dividing by $\sqrt{\left|\mathcal{N}\left(u\right)\right|\left|\mathcal{N}\left(v\right)\right|}$.

GCN learns node embeddings from a homogeneous graph with same node
types and relations. However, the pull request graph in Figure \ref{fig:ASTKG} is
a heterogeneous graph with different node types and different relation
types between them. In this case, inspired by RGCN \cite{schlichtkrull2017modeling},
for each node, we aggregate the feature information separately for
each type of relation. 

Mathematically it can be represented as follows:

\begin{equation}
h_{u}^{\left(k\right)}=\sigma\left(\sum_{r\in\mathcal{R}}\sum_{v\in\mathcal{N}_{r}\left(u\right)}\frac{W_{r}^{\left(k-1\right)}h_{v}^{\left(k-1\right)}}{\sqrt{\left|\mathcal{N}_{r}\left(u\right)\right|\left|\mathcal{N}_{r}\left(v\right)\right|}}+W_{0}^{\left(k-1\right)}h_{u}^{\left(k-1\right)}\right)\label{eq:rgcn}
\end{equation}

where $\mathcal{R}$ is the set of relations, $\mathcal{N}_{r}\left(u\right)$
is the set of neighbors of $u$ having relation $r$, $W_{r}^{\left(k\right)}$
is the relation-specific feature transformation weights for the $k^{\text{th}}$
layer; $W_{0}^{\left(k\right)}$ is the feature transformation weights
for the self node.

The set of relations $\mathcal{R}$ captures the semantic relatedness
of different types of nodes in the graph. This is generally determined
by the domain knowledge. For \nrr{} we identified
a bunch of useful relations as listed in Table \ref{tab:Relations-used}. 

\begin{table*}
\centering
\caption{\label{tab:Relations-used}Relations ($\mathcal{R}$) used for generating embeddings}
\begin{tabular}{cll}
\toprule
 & Relation & Semantic Description\tabularnewline
\midrule
1 & PullRequest - User & Captures the author or reviewer relationship between a pull request
and a user\tabularnewline

2 & PullRequest - File & Captures the related file modification needed for a pull request\tabularnewline

3 & PullRequest - Word & Captures the semantic description of a pull request through the words\tabularnewline

4 & File - Word & Captures the semantic description of a file.\tabularnewline

5 & Word - Word & Captures the related words in a window of size 5 in a sentence (in
the pull request title/description)\tabularnewline
\bottomrule
\end{tabular}

\end{table*}

In our experiments, we use a 2-layer GCN network, i.e., we set $k=2$ in Equation \ref{eq:rgcn}. 
With this, GCN can capture second order relations such as User-User,
File-File, User-File, User-Word, etc., which we believe are useful in capturing interesting
dependencies between various entities, such as related files, related
users, files authored/modified by users, words associated with users,  etc. While setting $k$ to a higher value can fold-in
longer distance relations, it is not clear if that helps or brings more
noise. We leave that exploration to our future work.

\subsection{Training the Model}

To learn the parameters of the model (i.e.,$W_{r}^{\left(\cdot\right)} $and $W_{0}^{\left(\cdot\right)}$)  we pose it as a \emph{Link Prediction}
problem. Here, we set the probability of existence of a link/edge
between two nodes $u$ and $v$ as proportional to the dot product
between their embeddings derived from the 2-layer GCN. In particular,
we set the link probability as equal to $\sigma\left(\mathbf{z}_{u}^{T}\mathbf{z}_{v}\right)$.
Here, $\sigma$ denotes the logistic function, and $\mathbf{z}_{u},\mathbf{z}_{v}$
denote the embeddings of nodes $u,v$ respectively (i.e., $\mathbf{z}_{u}=h_{u}^{\left(2\right)},\mathbf{z}_{v}=h_{v}^{\left(2\right)}$
from Equation \ref{eq:rgcn}). This probability is high when the nodes
$u$ and $v$ are connected in the graph. And, it is low when the
nodes $u$ and $v$ are not connected in the graph. Accordingly, we
prepare a training data set $\mathcal{D}$ containing records of triplets
$\left(u,v,y\right)$, where $\left(u,v\right)$ are the edges in
the graph and $y\in\left\{ 0,1\right\} $ denotes the presence or
absence of an edge between $u$ and $v$. Since there can be very
large number of node pairs $\left(u,v\right)$ where $u$ and $v$
are not connected, we employ random sampling to select a sizable number
of such pairs. The training objective is to minimize the cross-entropy
loss $\mathcal{L}$ in the Equation \ref{eq:loss}.

\begin{equation}
\mathcal{L}=-\frac{1}{\left|\mathcal{D}\right|}\sum_{\left(u,v,y\right)\in\mathcal{D}}y\text{log \ensuremath{\sigma\left(\mathbf{z}_{u}^{T}\mathbf{z}_{v}\right)}+\ensuremath{\left(1-y\right)}}\text{log}\left(\sigma\left(\mathbf{z}_{u}^{T}\mathbf{z}_{v}\right)\right)\label{eq:loss}
\end{equation}

Minimizing the above loss enforces the dot product of the embeddings
of the nodes $u,v$ to attain high value when they are connected by
an edge in the graph (i.e., when $y=1$), and a low dot product value
when they are not connected in the graph (i.e., when $y=0$). The
parameters of the model are updated as the training progresses to
minimize the above loss. We stop training when the loss function stops
decreasing (or the decrease becomes negligible).

\subsection{Inductive Recommendation for New Pull Requests}

GCN by design is a transductive model. That is, it can generate embeddings
only for  the nodes that are present in the graph during the training.
It cannot generate embeddings for the new nodes without adding those
nodes to the graph and retraining. On the other hand, inductive models
can infer embeddings for the new nodes that were unseen during the
training by applying the learned model to the new nodes. Since \NRR is a GCN-based model, we will not have embedding for the new pull request
$u^{\prime}$ at the inference time. We need to derive the embedding
for $u^{\prime}$ on-the-fly by applying Equation \ref{eq:rgcn}.
The challenge in deriving the embedding is in getting the correct
self embedding for $u^{\prime}$. That is, as per Equation \ref{eq:rgcn},
to generate $h_{u^{\prime}}^{\left(2\right)}$, we need trained $W_{0}^{\left(0\right)}$
and $W_{0}^{\left(1\right)}$, which are not available for the new
nodes. Hence we approximate the embedding of the new node by ignoring
its self embedding part in Equation \ref{eq:rgcn}, which leads to
the following approximation:

\begin{equation}
\mathbf{z}_{u^{\prime}}=h_{u^{\prime}}^{\left(2\right)}=\sigma\left(\sum_{r\in\mathcal{R^{\prime}}}\sum_{v\in\mathcal{N}_{r}\left(u^{\prime}\right)}\frac{W_{r}^{\left(1\right)}h_{v}^{\left(1\right)}}{\sqrt{\left|\mathcal{N}_{r}\left(u^{\prime}\right)\right|\left|\mathcal{N}_{r}\left(v\right)\right|}}\right)\label{eq:inference}
\end{equation}

Here, $\mathbf{z}_{u^{\prime}}$ is the embedding of the new pull
request $u^{\prime}$, $\mathcal{R^{\prime}}$ is the set of relations
involving the pull request node (i.e., PullRequest-User, PullRequest-File,
and PullRequest-Word), $W_{r}^{\left(1\right)}$ are the trained model
weights from the $2^{\text{nd}}$ layer of the GCN, and $h_{v}^{\left(1\right)}$
are the embeddings coming out of the first layer of the GCN. 

After obtaining the embedding of the new pull request as per Equation
\ref{eq:inference}, we can get the top $k$ reviewers for it by finding
the top $k$ closest users in the embedding space. That is,

\begin{equation}
\text{reviewers}_{k}\left(u^{\prime}\right)=\underset{v_{1}..v_{k}}{\text{argmax}}\left(\mathbf{z}_{u^{\prime}}^{T}\mathbf{z}_{v_{i}}\right)\label{eq:top_k_users}
\end{equation}

where $\mathbf{z}_{v_{i}}$ is the embedding of the user $v_{i}$.

Since our training objective enforces high $\mathbf{z}_{u}^{T}\mathbf{z}_{v}$
score when the likelihood of an edge $\left(u,v\right)$ is high,
 Equation \ref{eq:top_k_users} finds the users who are most likely
to be associated with the pull request, as reviewers. Finding top
$k$ reviewers in this way using their embeddings allows us to naturally
make use of complex relationships that are encoded in those embeddings
to capture user's relatedness to the pull request. 
\section{Experiments} \label{eval}
To assess the value of \NRR empircally, we pose three research questions:
\begin{enumerate}[font={\bfseries},label={RQ$_\arabic*$}, leftmargin=4\parindent]
\item \RQone
\item Under what circumstances does \NRR perform better than a \HM (and vice versa)?
\item What are developers' perceptions about \NRR? 
\end{enumerate}

The vast majority of code reviewer recommendation approaches are evaluated by comparing recommendations from the tool with historical code reviews and examining how often the recommended reviewers were the actual reviewers~\cite{dougan2019investigating}.
In line with this accepted practice, RQ1 asks how often the network is able to recommend the reviewers that the authors added.
However, as Dougan~\etal point out, there is an underlying (and often unstated) assumption that these are the correct reviewers\footnote{We would point out that if this assumption were correct, then there would be no need for a recommender in the first place!}~\cite{dougan2019investigating}.
To address this flawed assumption and pursue a more complete ground truth, we reach out to the reviewers recommended by \NRR that were \emph{not} recommended by a \hm. The results of this developer study help address RQ2 and RQ3.

For the purpose of conducting the experiments and comparative studies, we use a \hm{} built based on the heuristics proposed by Zanjani et al. \cite{zanjani2015automatically} which demonstrated that considering the history of source code files edited in a \pr{} in terms of authorship and reviewership is an effective way to recommend peer reviewers for a code change. This model is currently deployed in production at Microsoft. This gives us an opportunity to conduct comparative studies by observing the recommendations made by the \NRR{} and the telemetry generated from the production deployment.

\subsection{Methodology} \label{Methodology}

\subsubsection{Retrospective Evaluation}

To address RQ1, we construct a dataset of \EvalPullRequest{} code reviews, i.e. \pr{}--reviewer pairs, starting from \PRstartYear{} to evaluate \nrr{}. To keep training and validation cases separate, these nodes and their edges are not present in the graph during model training.
We use the following metrics, which are the most common measures for evaluating reviewer recommender approaches~\cite{thongtanunam2015should,ouni2016search,sulun2019reviewer,zanjani2015automatically,balachandran2013reducing,rahman2016correct}:

\MyPara{Accuracy}
We measure the percentage of \pr{}s from test data for which \nrr{} is able to recommend at least one reviewer correctly and report the percentage for top 1, 3, 5, and 7 reviewers suggested by the model.
Specifically, given a set of pull requests $P$, the top-k accuracy can be calculated using Equation~\ref{eq:topk}. The isCorrect($p$, Top-k) function returns value of 1 if at least one of top-k recommended reviewers actually review the pull request $p$; and returns value of 0 for otherwise.

\begin{equation}
\textrm{Top-}\mathit{k}\:\textrm{accuracy}(P)=\frac{\sum_{p\in P}^{} \textrm{isCorrect}(p, \textrm{Top-}\mathit{k})}{|P|}
\label{eq:topk}
\end{equation}

\MyPara{Mean reciprocal rank (MRR)} This metric is used extensively in recommender systems to assess whether the correct recommendation is made at the top of a ranked list \cite{10.5555/1394399}. MRR is calculated using Equation~\ref{eq:mrr}, where rank(candidates(p)) returns value of the first rank of correct reviewer in the recommendation list candidates(p).

\begin{equation}
\textrm{MRR}=\frac{1}{|P|}\sum_{p\in P}^{}\frac{1}{\textrm{rank(candidates(p))}}
\label{eq:mrr}
\end{equation}


\begin{table}
\centering
\caption{Pull request distribution across dimensions}
 \begin{tabular}{l|r} 
\toprule 
Repo size (\# of developers) & \# data \\
\midrule
Large & 220 \\
Medium & 200 \\
Small & 80 \\

\bottomrule
\end{tabular}
\label{tab:StartRandomSampling}
\end{table}

\subsubsection{User Study}
To address RQ2 and RQ3, we conduct a user study by reaching out to reviewers recommended by \NRR to see if they would be qualified to review the pull requests.

\MyPara{Sampling}
We select \UserStudySize{} recent \pr{}s from the test data set of 254K pull requests and randomly pick one of the top 2 recommendations by \NRR as the recommended reviewer to reach out. Note that \pr{}s selected  had not been recommended by the \hm{} and each recommended reviewer appears at most once.
The \pr{}s are collected from repositories having different number of developers using stratified random sampling following the distribution in Table~\ref{tab:StartRandomSampling}.  The categories are defined as follows:
number of developers: Large ( $> 100$ developers), Medium (between 25 and 100 developers), Small ($< 25$ developers).

\MyPara{Questionnaire} We perform the user study by posing a set of questions 
on what actions a reviewer might take when they were recommended for a specific \pr{}:
\begin{enumerate}
    \item Is this pull request relevant to you (as of the PR creation date, state)? \\
	A - Not relevant \\
	B - Relevant, I'd like to be informed about the pull request. \\
	C - Relevant, I'd take some action and/or I'd comment on the pull request.
	\item If possible, could you please explain the reason behind your choice?
\end{enumerate}

We avoid intruding in the actual work flow, yet still maintain an adequate level of realism by working with actual pull requests, thus balancing realism and control in our study \cite{10.1145/3241743}.
Note that, with \UseMacro{NumberTotalResponse} responses, this is one of the largest field studies conducted to understand the effects of an automated reviewer recommender system.

We divided the questionnaire among 4 people to conduct the user studies. The interviewers did not know these reviewers, nor had worked with them before. The teams working on the systems under study are organizationally far away from the interviewers. Therefore, they do not have any direct influence on the study participants. The interview format is semi-structured where users are free to bring up their own ideas and free to express their opinions about the recommendations.

We use the question (2) to collect user feedback and analyze it to generate insights about the perceptions of the developers about the automated reviewer recommendation systems (RQ$_3$). Namely, the factors that influence people to not lean towards using an automated reviewer recommendation system.

\subsubsection{Comparing with \HM{}}
To compare \NRR with the \hm{}, we select another \UserStudySize{} recent \pr{}s from the set of \pr{}s on which the \hm{} (currently deployed in production) has made recommendations, by following the same distribution as the \pr{}s selected for evaluating \nrr{} (Table \ref{tab:StartRandomSampling}). We then collect the recommendations made by the \hm{}, the subsequent actions performed by the recommended reviewers (changing the status of the \pr{}, or adding a code review comment, or both) for the selected \pr{}s from telemetry. The telemetry yields two benefits: 1. it helps us gather user feedback without doing another large-scale user study, as the telemetry captures the user actions already 2. it avoids the probable study participants from taking one more survey (and save time and frustration), because they already indicated their preferences on the \pr{} when it was active and when they were added as reviewers. 

An important point to keep in mind is, the \hm{} is adding recommended reviewers directly to the \pr{}s. This increases the probability of them taking an action even if they may not be an appropriate developer to conduct the review. The reason for this is that the reviewers are being selected and the assignment of them to the PR is public (everyone, including their managers, can see who is reviewing the \pr{}) \cite{rigby2011understanding}. If they do not respond, it might look like they are blocking the \pr{} progression. In contrast, \NRR's recommendations are validated through user studies, which are conducted in a private 1-1 setting where participants likely feel more comfortable indicating that they are not appropriate for the review. Reviewers can be open about their decisions in the user studies. Therefore, \nrr{} might be at a slight disadvantage.


\subsection{Results}
\subsubsection{\textbf{\RQone}}

To answer RQ$_1$, we examine who the \pr{} author invited to review a change and then check to see if \NRR recommended the same reviewers.
In this context, the ``correct'' recommendation is defined as the recommended reviewer being invited to the \pr.
While the author's actions may not actually reflect the ground truth of who is best able to review the change, most prior work in code reviewer recommendation evaluates recommenders in this way (see \cite{dougan2019investigating} for a thorough discussion of this) and so we follow suit here.
Table~\ref{tab:LinkPredictionAccuracy} shows the accuracy and MRR for \NRR across all 254K (pull request-reviewer) pairs.
In 73\% of the pull requests, \nrr{} is able to replicate the human authors' behavior in picking the reviewers in the top 3 recommendations which validates that \nrr{} matches the history of reviewer selection quite well. 
\footnote{Note that by design the \hm{} always includes the author-invited people to review, so we do not evaluate \hm{} in this approach.}


\begin{table}
\centering
\caption{Link prediction accuracy and MRR}
 \begin{tabular}{lrrrr} 
\toprule 
Metric & k = 1 & k=3 & k=5 & k=7 \\
\midrule
Accuracy & 0.50 & 0.73 & 0.78 & 0.80 \\
MRR & 0.49 & 0.61 & 0.68 & 0.72  \\
\bottomrule
\end{tabular}
\label{tab:LinkPredictionAccuracy}
\end{table}

\begin{table}
\centering
\caption{\TableCaptionUserStudyResults{}}
 \begin{tabular}{lll} 
\toprule
Repo size (\# of developers) & RM & \NRR \\
\midrule
Large & \UseMacro{LargeBaselineAcc} & \textbf{\UseMacro{LargeNRRAcc}}\\
Medium &  \UseMacro{MediumBaselineAcc}$^{\alpha}$ & \textbf{\UseMacro{MediumNRRAcc}}$^{\alpha}$\\
Small &  \textbf{\UseMacro{SmallBaselineAcc}}$^{\beta}$ & \UseMacro{SmallNRRAcc}$^{\beta}$\\
\bottomrule
\end{tabular}
\end{table}

\subsubsection{\textbf{RQ$_2$: Under what circumstances does \NRR perform better than a rule-based model (and vice versa)?}}
In Table \ref{tab:UserStudy}, we show the recommendation \engagementpct of \hm{} and \nrr{}. 
Specifically, on the sampled \UseMacro{User} data for the \nrr{} model and \hm{}, \engagementpct is calculated as the percentage of the recommended reviewers who are willing to engage in reviewing the pull requests. 
For \hm{}, reviewers who either change the status of the pull request or add a code review comment are considered as engaged. For \nrr{}, reviewers who say that the pull request is relevant and they would take some action are considered as engaged.

Generally, there is ``no model to rule them all". Neither of the models performs consistently better than the other in all the pull requests from repositories of all categories. As shown in table \ref{tab:UserStudy}, \nrr{} performs better on pull requests from large repositories and medium repositories while the \hm{} does well on pull requests from small repositories.  However, when we statistically tested for differences, Fisher exact tests~\cite{agresti2003categorical} only showed a statistically significant difference between the two approaches for large repositories ($p = 0.013$).

One observation that may explain this result is that due to their size, large software projects dominate the graph. Thus, \NRR is trained on many more pull requests from large projects than from smaller projects.  
If the mechanisms, factors, behaviors, etc., for reviewer selection are different in smaller projects than large ones, then the model is likely learn those used in larger projects.
This hypothesis could be confirmed by splitting the training data by project size and training multiple models.
However, as reviewer recommendation is most important in projects with many developers and that appears to be where \NRR excels, we do not pursue this line of inquiry.

We have observed that in small repositories usually with few developers, one or two experienced developers are more likely to take the responsibility of reviewing pull requests which accounts for the high accuracy of \hm{}.
However, this phenomenon in which a small number of experienced people in a particular repository are assigned the lion's share of reviews is problematic, and heuristics have been used to ``share the load''~\cite{asthana2019whodo}.
As the \NalandaGraph{} contains historical information about a developer across many repositories and PRs from different repositories may be semantically related, \NRR is able to leverage more information per developer and per PR, which may avoid this problem.

The following feedback received from the user study (question (2)) also demonstrates that \NRR identifies relevant and qualified reviewers who traditional reviewer recommenders miss:

\anecdote{This PR is created in a repository on which our service has a dependency on. I would love to review these PRs. In fact, I am thinking of asking \textbf{x} on these PRs going forward.}

\anecdote{I never reviewed \textbf{y}’s PRs. I work with her on the same project and know what she is doing. I am happy to provide any feedback (of course if she’d like :))}

\anecdote{The content of the PR might impact another repository that I have ownership of because we use some of the components in that lib. Based on that I would say it is a relevant PR and I will not mind reviewing it.}

\subsubsection{\textbf{RQ$_3$: What are developer's perceptions about an automated reviewer recommendation model?}}
\begin{table}
\centering
\caption{\TableCaptionUserResponse{}}
 \begin{tabular}{ll} 
\toprule
Category & \# of responses (\%) \\
\midrule
I will review this \pr{} & \UseMacro{NumberResponseC} (\UseMacro{RateResponseC}) \\
I'd like to be added to this \pr{} & \UseMacro{NumberResponseB} (\UseMacro{RateResponseB}) \\
This \pr{} is not relevant to me & \UseMacro{NumberResponseA} (\UseMacro{RateResponseA}) \\
\bottomrule
\end{tabular}
\end{table}

We show the distribution of user study responses in Table ~\ref{tab:UserResponse}. Out of \UseMacro{UserDataSampleSize} user study messages we sent, \UseMacro{NumberTotalResponse} users responded.
\UseMacro{RatePositiveResponse} of the users give positive feedback saying that the given pull request is relevant to them to some degree. In this, 
\UseMacro{RateResponseB} of the users say they would like to be informed about the given pull request. \UseMacro{RateResponseC} of the users say that they would like to take some action and/or leave comment on the pull request.
\UseMacro{RateNegativeResponse} of the users give negative feedback saying that the given pull request is not relevant.

\begin{table*}
 \centering
\caption{\TableCaptionNegativeFeedback{}}
 \begin{tabular}{llr} 
\toprule
Category & Feedback & \# of feedback (\%)\\
\midrule
\feedback{1} & This \pr{} is no longer relevant to me & \UseMacro{NumberFeedbackCategory1Combined} (\UseMacro{RateFeedbackCategory1Combined})\\
\feedback{2} & Never participate in code review & \UseMacro{NumberFeedback3} (\UseMacro{RateFeedback3}) \\
\feedback{3} & Pull request does not need reviewer & \UseMacro{NumberFeedback4} (\UseMacro{RateFeedback4}) \\
\bottomrule
\end{tabular}
\end{table*}


To understand the reason that users do not like \NRR's recommendations, we analyze the  negative feedback (comments/anecdotes from the developers) and classify them into \NegFeedbackCats{} categories with their distribution shown in Table~\ref{tab:NegFeedback}.
To offer an impression, we show some typical negative quotes that we received from users.

\UseMacro{RateFeedbackCategory1Combined} of the negative feedback we received said that the pull request is no longer relevant to them and \UseMacro{RateFeedback1} of them said it is because they started to work in a different area and
\UseMacro{RateFeedback2} of them mentioned that they do not work in this repository because of switching groups or transferring teams:
\anecdote{Not relevant since I no longer work on the team that manages this service.}
\UseMacro{RateFeedback3} of the users mentioned that they are actually never involved in code review:
\anecdote{I'm a PM. I'm less interested in PR in general. Only when I'm needed by the devs and then they mention me there.}
Two users said that the pull requests we provided does not need to be reviewed:
\anecdote{Let me explain. This is an automated commit that updates the version number of the product as part of the nightly build. It pretty much happens every night. So it doesn't need reviewer like a traditional pull request would.}

From users' negative feedback, we learn that in order to improve \nrr{}
we need to include several extra factors. First, our \NalandaGraph{} should take the people movement into consideration and update the graph dynamically, namely identifying inactive users and removing edges or decaying the weight on the edges between user nodes and repository nodes.

Second, \nrr{} should include and learn the job role for every user in the \NalandaGraph{} through node embeddings, such as SDE, PM, so that it can filter those irrelevant users and suggest the reviewers more precisely. 

Third, before running the \nrr{}, some heuristic rules can be designed to filter the automatic, deprecated pull requests.

Besides the negative feedback, we receive a lot of credits from users:

\anecdote{The recommendation makes a lot of sense since I primarily contributed to that repository for a few years. However, a recent re-org means I no longer work on that repository.}

\anecdote{I am lead of this area and would like to review these kinds of PRs which are likely fixing some regressions.}

They validate our claim that \nrr{} does consider the interactions between users and files, and the recommendations are understandable by humans.
Since \nrr{} is trained and evaluated on historical pull requests starting from 
\PRstartYear{}, it is hard to reconstruct the situation where the pull requests were created and many users complain that it is difficult to recall the context of the pull requests, thus putting \nrr{} in a disadvantage. We expect it will have better performance in the actual production.

\begin{table*}[t]
\centering
\caption{Link prediction accuracy and MRR for various configurations of parameters}
  \begin{tabular}{lrrrrrrrr}
    \toprule
    \multirow{2}{*}{Models} &
    \multicolumn{4}{c}{Accuracy} &
      \multicolumn{4}{c}{MRR} \\
          \cmidrule(lr){2-5}
            \cmidrule(lr){6-9}          
    &  k = 1 & k=3 & k=5 & k=7 & k = 1 & k=3 & k=5 & k=7 \\
    \hline
    (1) No words or files & 0.02 & 0.08 & 0.13 & 0.16 & 0.01 & 0.04 & 0.05 & 0.06 \\

    (2) Words only & 0.21 & 0.30 & 0.32 & 0.34 & 0.21 & 0.25 & 0.26 & 0.32 \\
    
    (3) Files only & 0.29 & 0.69 & 0.73 & 0.76 & 0.29 & 0.48 & 0.49 & 0.50 \\
    (4) Words + Files & \textbf{0.49} & \textbf{0.73} & \textbf{0.77} & \textbf{0.80} & \textbf{0.49} & \textbf{0.61} & \textbf{0.68} & \textbf{0.72} \\
    \bottomrule
  \end{tabular}
  \label{tab:AblationStudy}
\end{table*}

\subsubsection{Ablation  Study} 

To evaluate the contribution of each of the entities in \NRR, we perform an ablation study, with results shown in Table~\ref{tab:AblationStudy}. Specifically, we first remove the entities from the \NalandaGraph{} and training data, and then retrain the graph convolutional neural network. We find that ablating each entity deteriorates performance across metrics. After removing word entities and file entities from graph, i.e. the \NalandaGraph{} only contains user and pull requests entities, the model can hardly recommend correct reviewers.
By comparing (1) and (2), (1) and (3), we demonstrate the importance of semantic information and file change history introduced by file entities in recommending reviewers and file entities give more value than words. 
Looking at (3) and (4), we observe boost in performance when adding semantics information on top of the file change and review activities, which underlines our claim that incorporating information around interactions between code contributors as well as the semantics of code changes and their descriptions can help identify the best reviewers.

\section{Threats and Limitations}

As part of our study, we reached out to people who were not invited to a review but that \NRR recommended as potential reviewers.
It is possible that their responses to our solicitations differed from what they may have actually done if they were unaware that their actions/responses were being observed (the so-called Hawthorne Effect~\cite{adair1984hawthorne}).
Microsoft has tens of thousands of developers and we were careful not to include any repositories or participants that we have interacted with before or might have a conflict of interest with us.
Nonetheless, there is a chance that respondents may be positive about the system because they wanted to make the interviewers happy.

The \kg{} contains information about \emph{who} was added as a reviewer on a PR, but it does not explain \emph{why} that person was added or if they were added as the result of a reviewer recommendation tool.
Thus, in our evaluation of how well \NRR is able to recommend reviewers that were historically added to reviews, it is unclear how much of history comes from the \hm{} recommender and how much from authors without the aid of a recommender.

When looking at repository history, the initial recommendation by the \hm{} is based on files involved in the initial review, while \NRR includes files and descriptions in the review's final state.  If the description or the set of files was modified, then \NRR may have a different set of information available to it than it would have had it been used at the time of PR creation. 

In our evaluation of \NRR, we use a training set of PRs to train the model and keep a hold out set for evaluation.  
These datasets are disjoint, but they are not temporally divided.
In an ideal setting all training PRs would precede all evaluation PRs in time and we would evaluate our approach by looking at \NRR's recommendation for the the next unseen PR (ordered by time), then add that PR to the \kg, and then retrain the model on the updated graph for the following PR and repeat until all PRs in the evaluation set were exhausted.
This form of evaluation proved too costly and time consuming to conduct and so we used a random split of training and testing data sets.

We sampled the 500 PRs from the population using a random selection approach.  We selected sample size in an effort to avoid bias and confounding factors in the sample, but we cannot guarantee that this data set is free from noise, bias, etc.

\section{Future Work} \label{FutureWork}
In this work we showed that a simple GCN style model is able to capture
complex interaction patterns between various entities in the code
review ecosystem and can be used to predict relevant reviewers for
pull requests effectively. While this method is very promising on
large sized repositories, we believe that the method can be improved
to make good recommendations on other repositories too by training
repository type specific models. In this work we mainly focused on
using interaction graph of various entities (pull requests, users,
files, words, etc.) to learn complex features through embeddings.
We neither captured any node specific features (e.g., user-specific features,
file-specific features, etc.) nor any edge specific features (e.g., how long
ago user authored/modified files, whether two users belong to the same
org or not, etc.). Incorporating such features may help the model
learn even complex patterns from the data and further improve the
recommendation accuracy. Furthermore, we believe that a detailed study
of effect of model hyper-parameters (such as embedding dimension,
number of GCN layers, different activation functions, etc.) on the
recommendation accuracy will be a very useful result. We intend to
explore these directions in our future work.

The techniques explained in this paper and the \nrr{} system are generic enough to be applied on any dataset that follows a GIT based development model. Therefore, we see opportunities for implementing \nrr{} for source control systems like GitHub and GitLab.
\section{Conclusion}
In this work, we seek to to leverage additional recorded information in software repositories to improve reviewer recommendation and address the weakness of the approaches that rely only on the historical information of changes and reviews.

To that end we propose \nrr{}, a novel Graph-based machining learning model that leverages a socio-technical graph built from the rich set of entities (developers, repositories, files, pull requests, work items, etc.) and their relationships in modern source code management systems. We train a Graph Convolutional Neural network (GCN) on this graph to learn to recommend code reviewers for \pr{}s. 

Our retrospective results show that in 73\% of the \pr{}s, \nrr{} is able to replicate the human \pr{} authors' behavior in top 3 recommendations and it performs better than the \hm{} in production on \pr{}s in large repositories by 94.7\%. A large-scale user study with 500 developers showed 67.6\% positive feedback, and relevance in suggesting the correct code reviewers for \pr{}s.

Our results open new possibilities for incorporating the rich set of information available in software repositories and the interactions that exist between various actors and entities to develop code reviewer recommendation models. We believe the techniques and the system has a wider applicability ranging from individual organizations to large open source projects. Beyond code reviewer recommendation, future research could also target other recommendation scenarios in source code repositories that could aid software developers leveraging the \kg{}s.
\section{Data Availability}
We are unfortunately unable to make the data involved in this study publicly available as it contains personally identifiable information as well as confidential information.
Access to the data for this study was made under condition of confidentiality from Microsoft
and we cannot share it while remaining compliant with the  General Data Protection Regulation (GDPR)~\cite{gdpr}.

\balance
\bibliographystyle{IEEEtran}
\bibliography{Paper}

\end{document}